\documentclass[9pt,conference]{article}
\usepackage{graphicx}
\usepackage{subfigure}
\usepackage[small]{caption}
\usepackage{setspace}
\usepackage{times}
\usepackage{amsmath}
\usepackage{amssymb}
\usepackage{acronym}
\usepackage{multirow}
\usepackage{balance}
\usepackage{cite}
\usepackage{authblk}
\begin{document}
\title{HoLiSwap: Reducing Wire Energy in L1 Caches}
\date{}

\author[1]{Yatish Turakhia}
\author[2]{Subhasis Das}
\author[3]{Tor M. Aamodt}
\author[4]{William J. Dally}

\affil[1,2,4]{Department of Electrical Engineering, Stanford University}
\affil[3]{Department of Electrical and Computer Engineering, University of British Columbia}
\affil[1,2,4]{\textit{\{yatisht, subhasis, dally\}@stanford.edu   $^3$aamodt@ece.ubc.ca}}
\maketitle

\begin{abstract}
  This paper describes HoLiSwap a method to reduce L1 cache wire energy, a significant fraction of total cache energy, by
swapping {\em hot} lines to the cache way nearest to the processor. 
We observe that (i) a small fraction (\textless3\%) of cache lines ({\em hot} lines) serve over 60\% of the L1 cache accesses and (ii) the difference in wire energy between the nearest and farthest cache
subarray can be over 6$\times$. 
Our method exploits this difference in wire energy to dynamically identify {\em hot} lines and swap them to the nearest physical way in a set-associative L1 cache. 
This provides up to 44\% improvement in the wire energy (1.82\% saving in overall system energy) with no impact on the cache miss rate and 0.13\% performance drop. 
We also show that HoLiSwap can simplify way-prediction. 
\end{abstract}

%

\section{Introduction}\label{sec:introduction}\par

%
%

Data movement across the memory hierarchy is increasingly becoming the dominant component of energy, both in smartphones and supercomputers. L1 caches expend up to half of the total movement energy~\cite{pandiyan2014quantifying, kestor2013quantifying} with roughly equal distribution in instruction (L1-I) and data (L1-D) cache.  
%
%

In this paper, we propose HoLiSwap\footnote{Used as an acronym for Hot Line Swap.}, an organization that migrates {\em hot} cache lines to the nearest way of a set associative cache to  minimize L1 cache wire energy with negligible impact performance or area. 

%
%
\begin{figure}[htp]
\vspace{-15pt}
\centering
\includegraphics[width=0.8\textwidth]{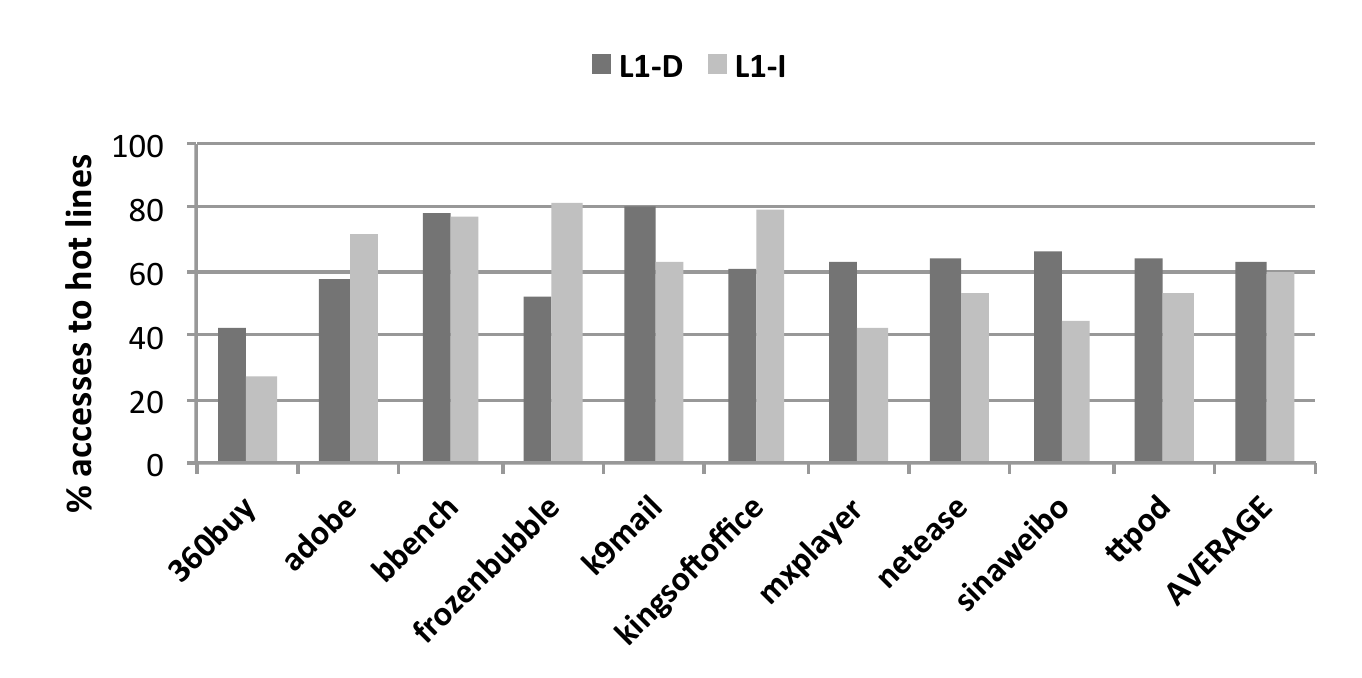}
\vspace{-15pt}
\caption{Percentage of total accesses to hot cache lines in L1 cache.
}
\label{fi:access}
\end{figure}

\begin{figure}[htp]
\vspace{-15pt}
\centering
\includegraphics[width=0.8\textwidth]{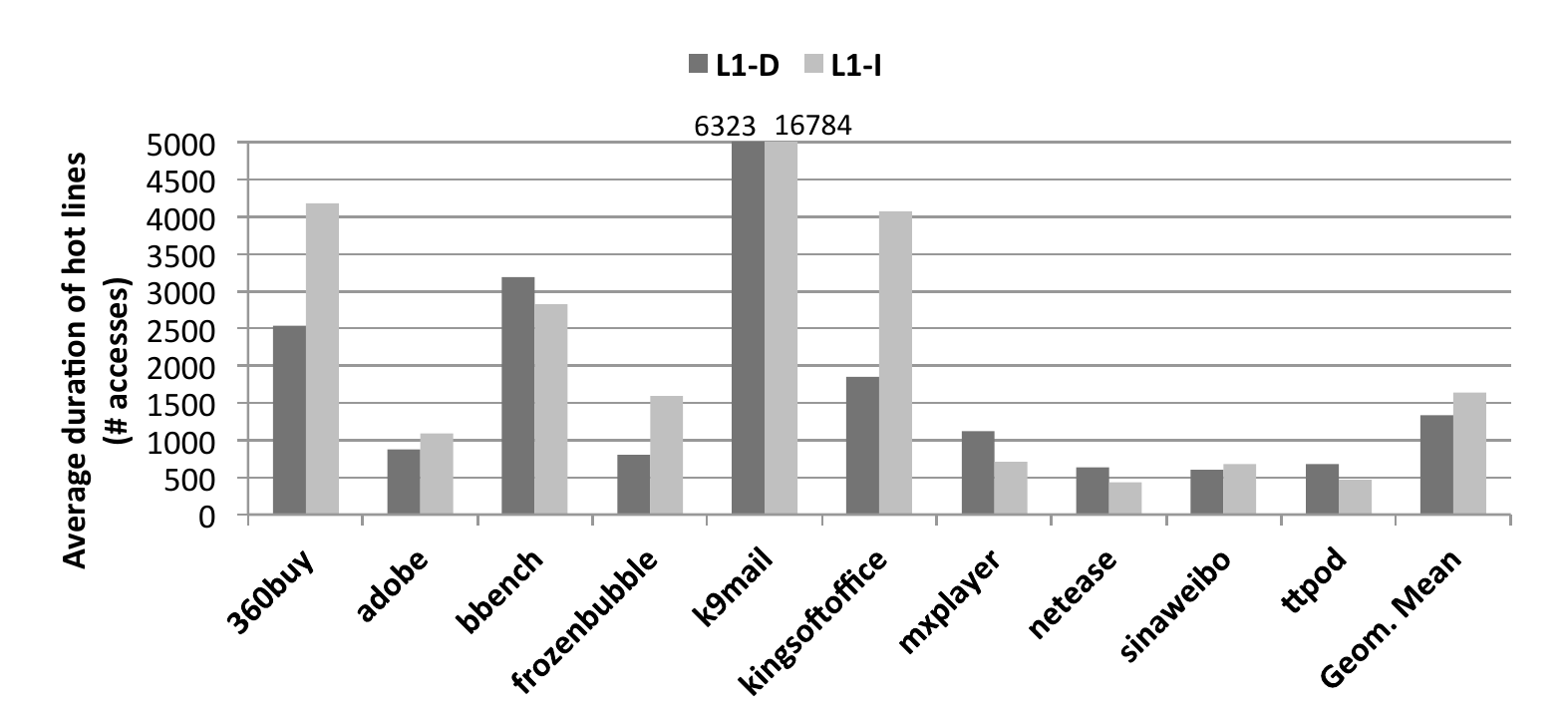}
\vspace{-15pt}
\caption{Average duration (\# accesses to its set) of hot cache lines.}
\label{fi:life}
\end{figure}

HoLiSwap is motivated by two observations:
First, certain cache lines are {\em hot} in that they have a long cache lifetime and are frequently accessed during this lifetime. Figure~\ref{fi:access} shows the percentage of accesses to hot lines (lines with $\ge$64 hits per 128 set accesses) in L1 caches in the Moby benchmarks~\cite{huangmoby}. 
The figure shows that about 60\% of all accesses are to hot lines.
Figure~\ref{fi:life} shows that the average hot line remains active for 1,341 set accesses in L1-D and 1636 set accesses for L1-I cache (compared to 125 accesses for a non-hot line).   On these
benchmarks only 2.88\% (2.45\%) of all lines in L1-D (L1-I) cache are hot.

Second, the wire energy for the cache way nearest the processor can be as much as 6$\times$ lower than the wire energy for the farthest way. Thus, HoLiSwap migrates the hot cache lines to the nearer cache ways, leveraging this asymmetry in energy consumption of cache ways. As we show in Section~\ref{sec:results}, this strategy can reduce L1 cache wire energy consumption by 45\% on an average for the Moby benchmark suite.


Prior work on L1 cache energy has focused on minimizing SRAM access energy. 
For instance, way-prediction~\cite{powell2001reducing}  speculatively predicts and accesses a single way for load instructions in parallel with tag lookup, resulting in lower access energy for correct predictions. 
Filter cache~\cite{ref:filtercache}, or L0 cache, adds a small cache between the processor and the  L1-cache to reduce L1-cache accesses. 
A number of techniques have been proposed to minimize the latency and energy overhead incurred on a filter cache miss~\cite{bardizbanyan_designing_2013, lee_filter_2014}.

HoLiSwap is similar in spirit to NuRAPID~\cite{chishti2003distance}, which relocates frequently accessed cache lines to faster subarrays
to improve performance of  last-level NUCA caches that employ sequential look-up of tag and data arrays. HoLiSwap improves upon NuRAPID and NUCA by first identifying  hot lines rather than indiscriminately swapping the most recently used line after every reference, leading to lower swapping overheads, 	both in energy and performance. We show in Section~\ref{sec:results} that frequent swaps in L1 cache could have high performance overhead. Also, L1 caches, in contrast to NUCA, have lower latency variation between subarrays and a substantially larger performance implication with cache port blocking. HoLiSwap focuses on minimizing wire energy rather than latency, and may use more complex lookup mechanisms, such as parallel lookup or way-prediction, as described in Section~\ref{sec:lookup}.

\section{Description}
\label{sec:description}


\begin{figure}
\vspace{-10pt}
\centering
\includegraphics[width=0.8\textwidth]{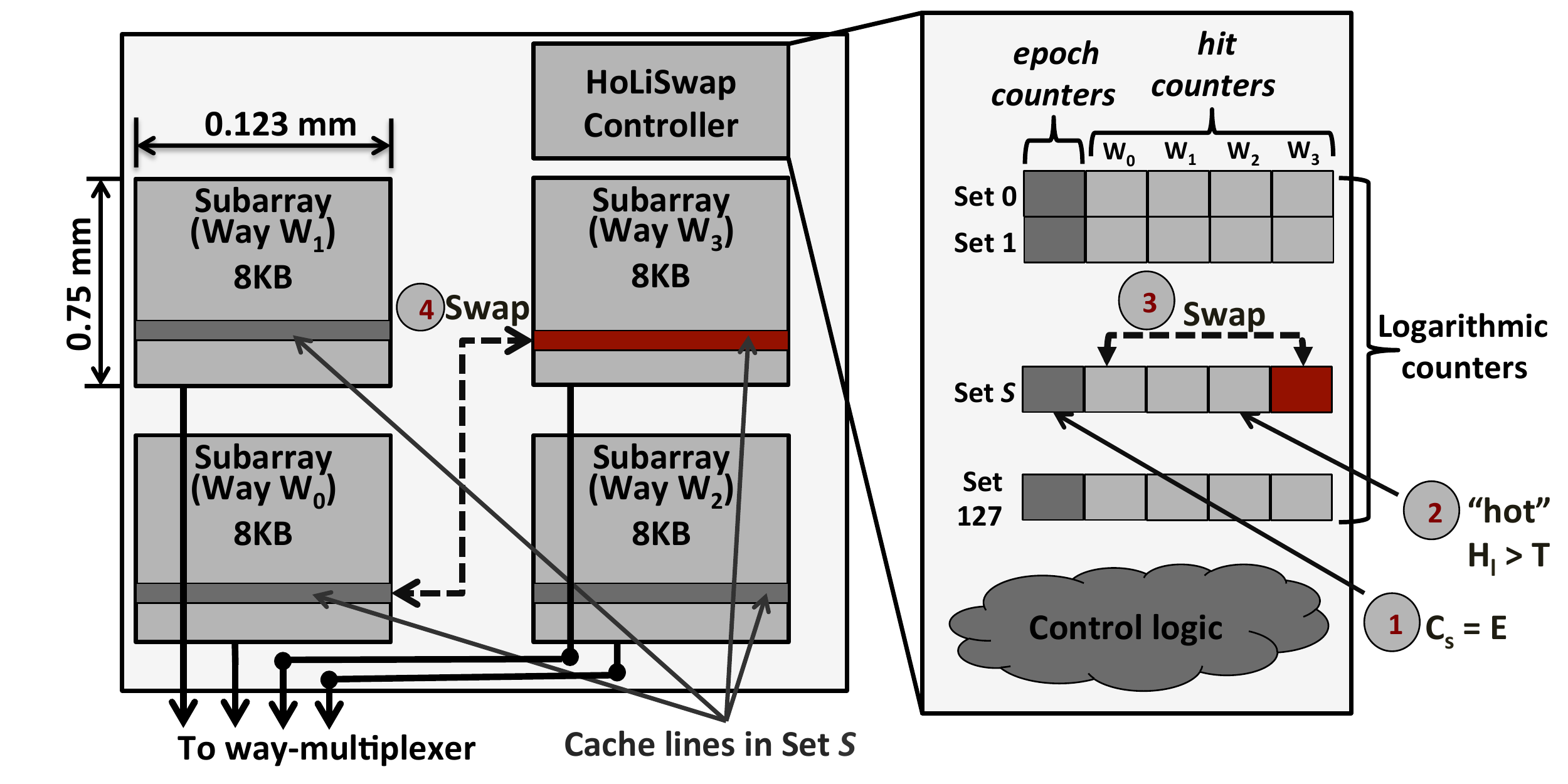}
\vspace{-10pt}
\caption{Overview of HoLiSwap.
}
\label{fi:overview}
\end{figure}
%


Figure~\ref{fi:overview} shows the organization of  32KB, 4-way set associative cache using HoLiSwap.  Each way is stored in a separate 8KB subarray.  To detect  hot cache lines, the controller maintains an epoch counter $C_s$ for each set $s$ and a hit counter $H_l$ for each line $l$.   At the start of each epoch, all counters are zero.  $C_s$ is incremented on each access to $s$ and $H_l$ is incremented on each hit to line $l$.  When $C_s = E$ a new epoch is started and all counters in the set are zeroed.  If at any time $H_l \ge T$ (where $T$ is a threshold), line $l$ is considered {\em hot} and the {\em hottest} line (most accesses) is swapped to way $W_0$, the lowest energy way. The ports of the cache are blocked when the swap is performed,  incurring a small performance penalty. 

To reduce the overhead of maintaining  counters , HoLiSwap uses logarithmic counters to store only the exponent (base 2) of $H_l$ and $C_s$.  At all times after the initial increment from 0 to 1, $H_l = 2^e$.  On a hit to line $l$ the exponent $e$ of  $H_l$ is incremented with probability $1/2^e$.   This representation reduces the storage overhead by 50\% for the 32KB 4-way set associative cache (Section~\ref{sec:methodology}) with little drop in  energy saving (\textless 2\%) compared to an exact count.
%
%



\label{sec:lookup}



HoLiSwap uses tri-state buffers to gate the data lines from the subarrays to the processor based on the
tag comparison to
prevent the unselected lines from toggling and dissipating energy.
This gating is possible because latency to access the smaller tag array is lower than latency to access the larger data subarrays (0.28ns for 2KB tag array vs. 0.5ns for 8KB data subarray).
%
%
%
%
Energy is still consumed cycling the bit lines of the unselected ways.

The HoLiSwap can optionally employ way prediction~\cite{powell2001reducing}  to avoid cycling the unselected data subarrays --- saving further energy, but with a small penalty in performance due to mispredictions.
Because the HoLiSwap migration strategy results in most accesses being made to $W_0$, always predicting $W_0$ provides a reasonably accurate way-predictor (as described in Section~\ref{sec:results}) at a much lower implementation cost.

%



\section{Methodology}
\label{sec:methodology}

\begin{table}
 \centering
\caption{Total energy (in pJ) to access different ways of a 4-way 32KB L1 cache for sequential and parallel lookup mechanisms. Wire energies are indicated in brackets.}    
\resizebox{0.8\textwidth}{!}
{
\begin{tabular}{|r|l|l|l|l|}
    \hline
 & Way 0 & Way 1 & Way 2 & Way 3 \\
    \hline
   Sequential & 5.7 (1.6) & 8.8 (4.7) & 10.9 (6.8) & 14.0 (9.9)\\
   Parallel & 18.0 (1.6) & 21.1 (4.7) & 23.2 (6.8) & 26.3 (9.9)\\
    \hline
    \end{tabular}}
    
    \label{tab:energy_values}
\end{table}

We evaluate HoLiSwap using  Gem5~\cite{binkert2011gem5} to simulate a 3-way out-of-order ARM processor. 
We use 10 Android applications from the Moby benchmark suite~\cite{huangmoby} representing popular and emerging mobile workloads.
We simulate 10M instructions (which typically contains 1-5M load/store instructions) for each benchmark from checkpoints collected at every 100M instructions. 
%
%

HSPICE is used to model the energy of a 32KB 4-way set associative L1 cache with the physical organization shown in Figure~\ref{fi:overview}. The output wires from the subarrays are routed to the way-multiplexer near way $W_{0}$ using Manhattan routing. We use PTM wire and CMOS models~\cite{ref:ptm} at the 22nm technology node.
The model includes output wire energy and  SRAM array access energy (including the decoder and sense amp energy). 
%
%
Table~\ref{tab:energy_values} shows the total energy (SRAM + wire) and the wire energy required to access each way of the L1 cache for sequential and
parallel lookup. 
For sequential lookup, we model the sum of array access energy and  output wire energy of the accessed way (subarray).
For parallel lookup, we model the sum of the array energies of all ways and the output wire energy of the required way (since remaining output wires have been gated). 
The parallel lookup mechanism offers a narrower energy range for different ways since the total energy is largely dominated by the parallel access to all ways. There is over 6$\times$ variation in wire energy from closest to farthest way.
%
%


We set $E=256$ and $T=128$.  These settings provided the most gain experimentally (in terms of energy-delay product) over a sweep of $E$ from 8 to 1024 accesses. Energy improvements are relatively insensitive to $T$, but setting $T$ to $E/2$ helps in leveraging logarithmic counters and ensures that at most one line is hot. The cache is blocked for 4-cycles on every swap and the energy overhead (2 reads, 2 writes) for each swap is accounted for. 
Storing the 4-bit exponential counters ($C_s$ and $H_l$) for each set
requires 20-bits of overhead (instead of 40-bits for exact count). 
We also account for the overhead of the simple controller.

We evaluate the benefits of HoLiSwap migration strategy on four different L1-D cache designs: (1) {\sc Sequential}: tags are accessed first
and then only the selected data arrays are accessed --- requiring 3-cycle hit latency; (2) {\sc Parallel}: tag and data arrays area accessed in parallel which reduces the hit latency to 2-cycles; (3) {\sc Prediction}: We use a PC-based way-predictor based on~\cite{powell2001reducing} without HoLiSwap migration strategy as baseline and compare it with the static way-predictor having HoLiSwap migration strategy as described in Section~\ref{sec:lookup}; (4) {\sc Filter}: This design uses a 1KB single cycle directly-mapped L0 cache in addition to the 32KB L1-D cache. The access energy of L0 cache is 4.1 pJ. 

For {\sc Prediction}, A 2-cycle hit latency is incurred for correctly predicted loads  and an additional cycle for all stores  and incorrectly predicted loads. Access energy to each cache line corresponds to the sequential mechanism as only one way is probed at a time. We also disable selective direct-mapping since it affects the miss rate of the cache. For performance reasons, L1-I cache typically uses {\sc Parallel} scheme only.


\section{Results}
\label{sec:results}

%
%

    

\begin{figure}
\vspace{-10pt}
\centering
\includegraphics[width=0.8\textwidth]{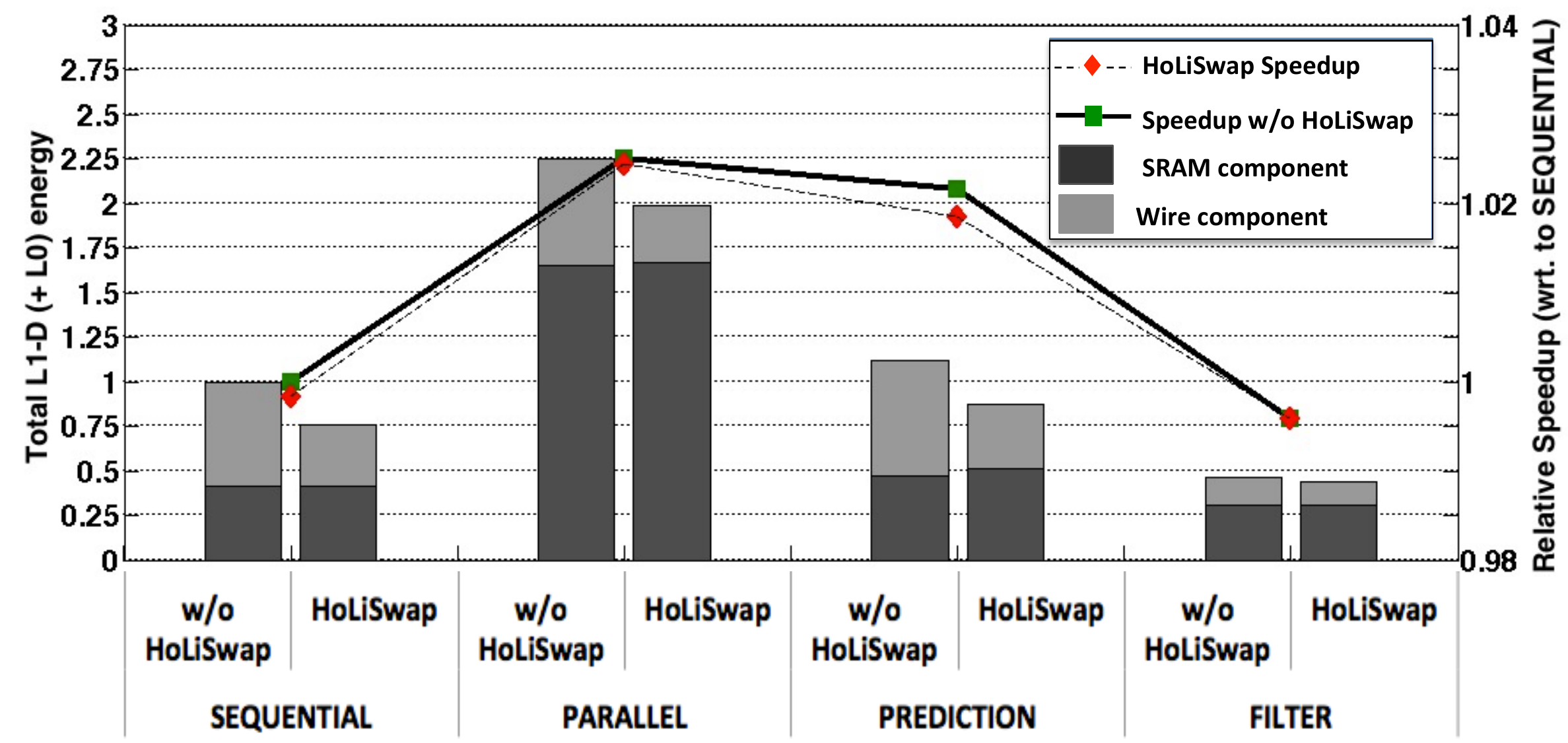}
\vspace{-10pt}
\caption{Energy-Performance (normalized to {\sc sequential}) with and without HoLiSwap migration in L1-D cache. 
}
\label{fi:energy_perf}
\end{figure}

\begin{figure}
\vspace{-10pt}
\centering
\includegraphics[width=0.8\textwidth]{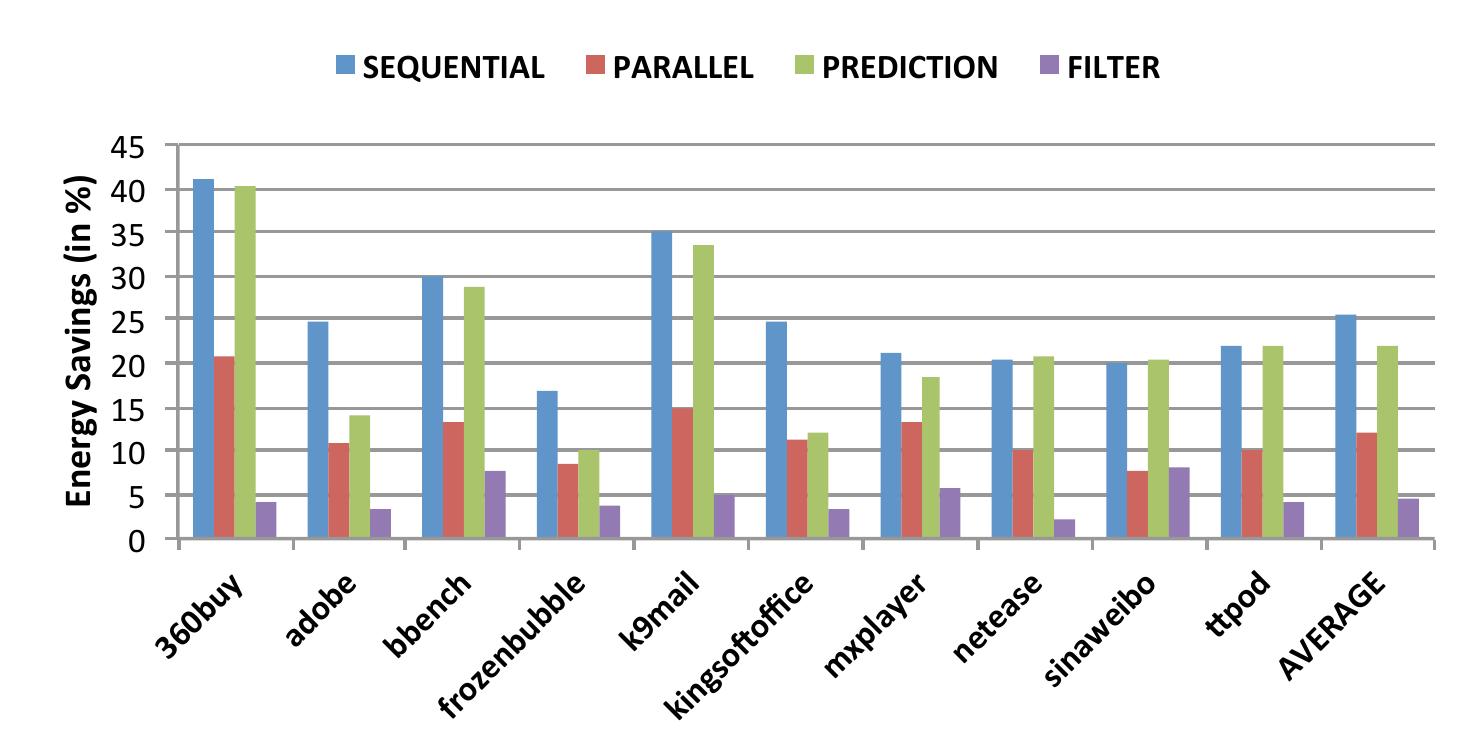}
\vspace{-10pt}
\caption{Energy savings in L1-D cache with the HoLiSwap migration policy for the four cache designs.
}
\label{fi:energy_savings}
\end{figure}

\begin{figure}
\vspace{-10pt}
\centering
\includegraphics[width=0.8\textwidth]{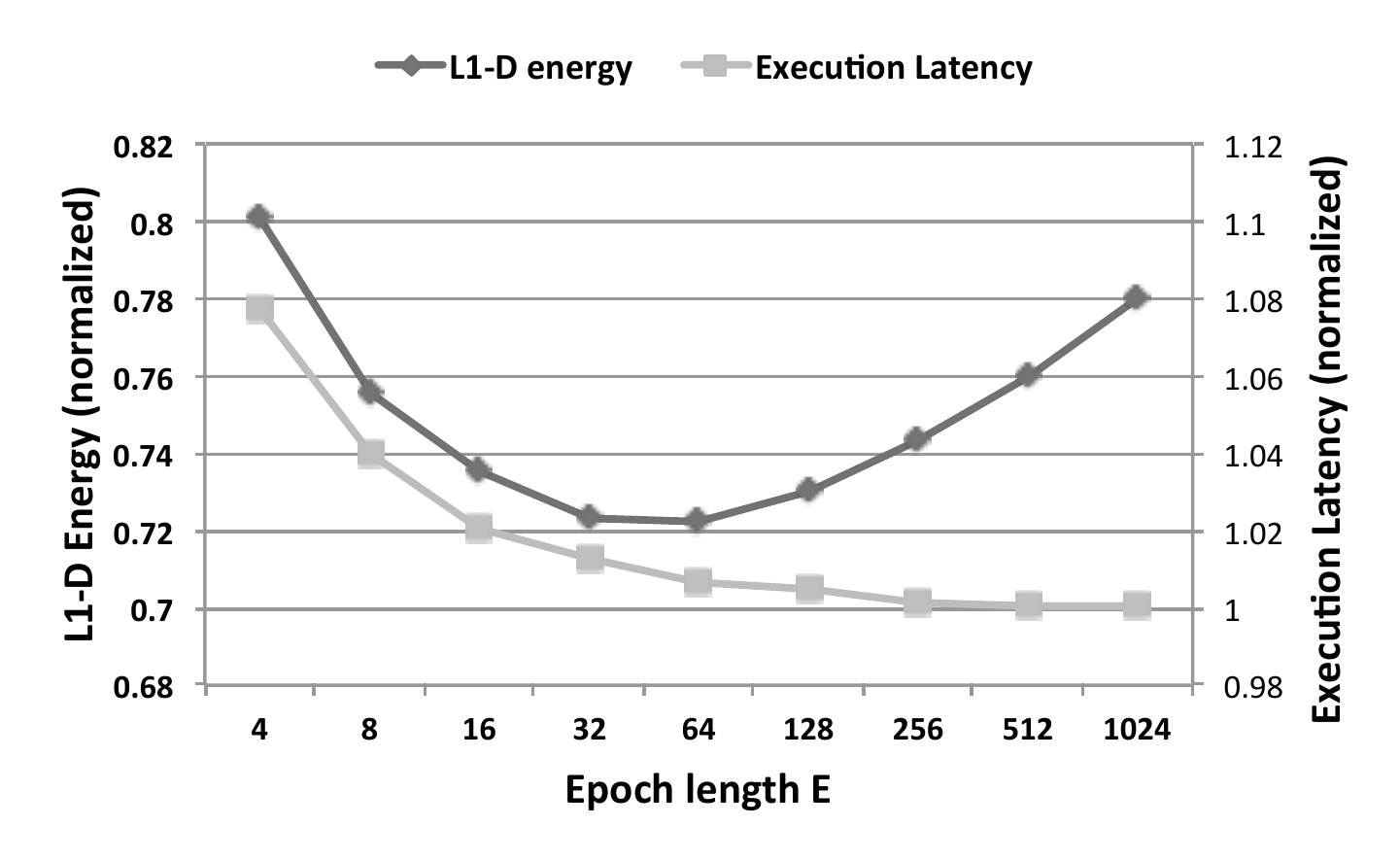}
\vspace{-10pt}
\caption{Normalized L1-D cache energy and execution latency vs. epoch length E.
}
\label{fi:sweep}
\end{figure}

\begin{figure}
\vspace{-10pt}
\centering
\includegraphics[width=0.9\textwidth]{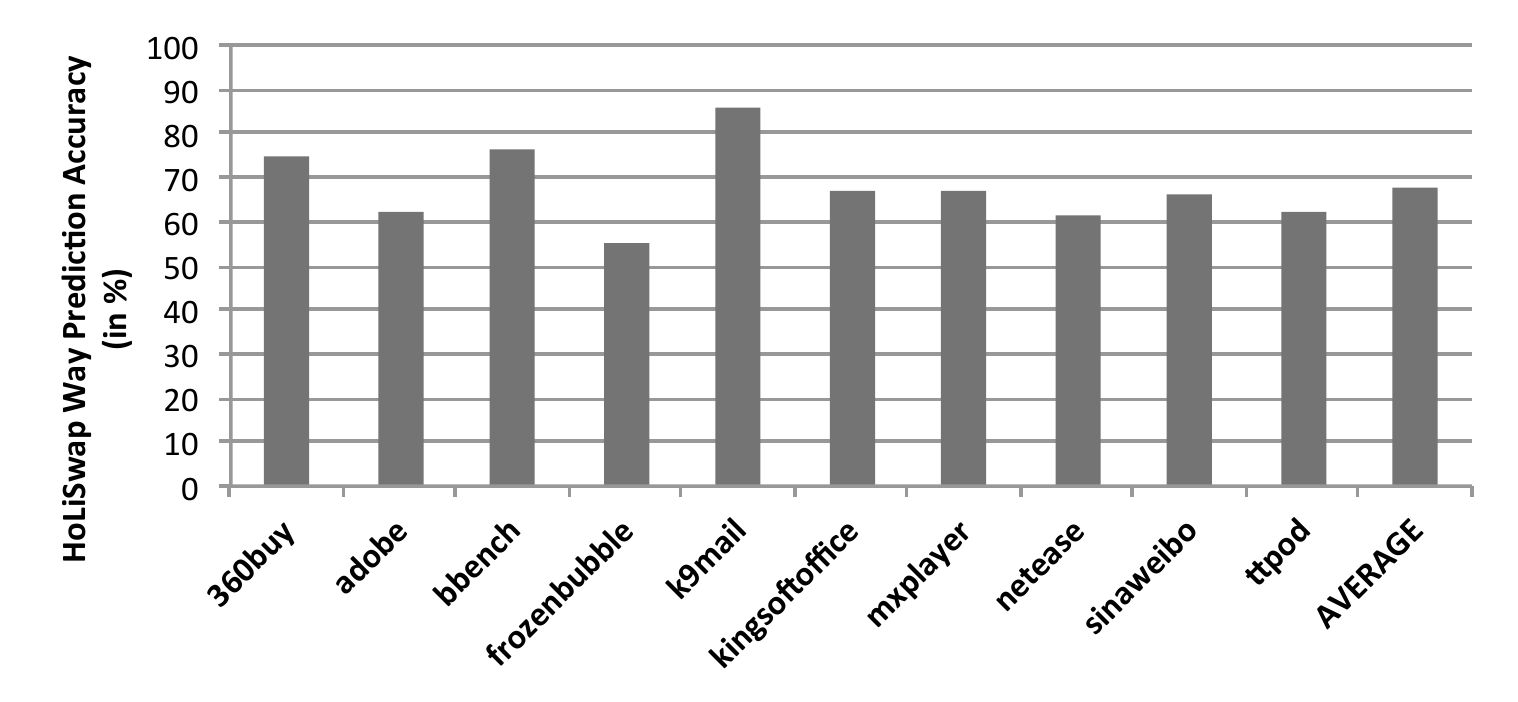}
\vspace{-10pt}
\caption{Way prediction accuracy for the static way-predictor in HoLiSwap.}
\label{fi:way_pred}
\end{figure}

\begin{figure}
\vspace{-10pt}
\centering
\includegraphics[width=0.9\textwidth]{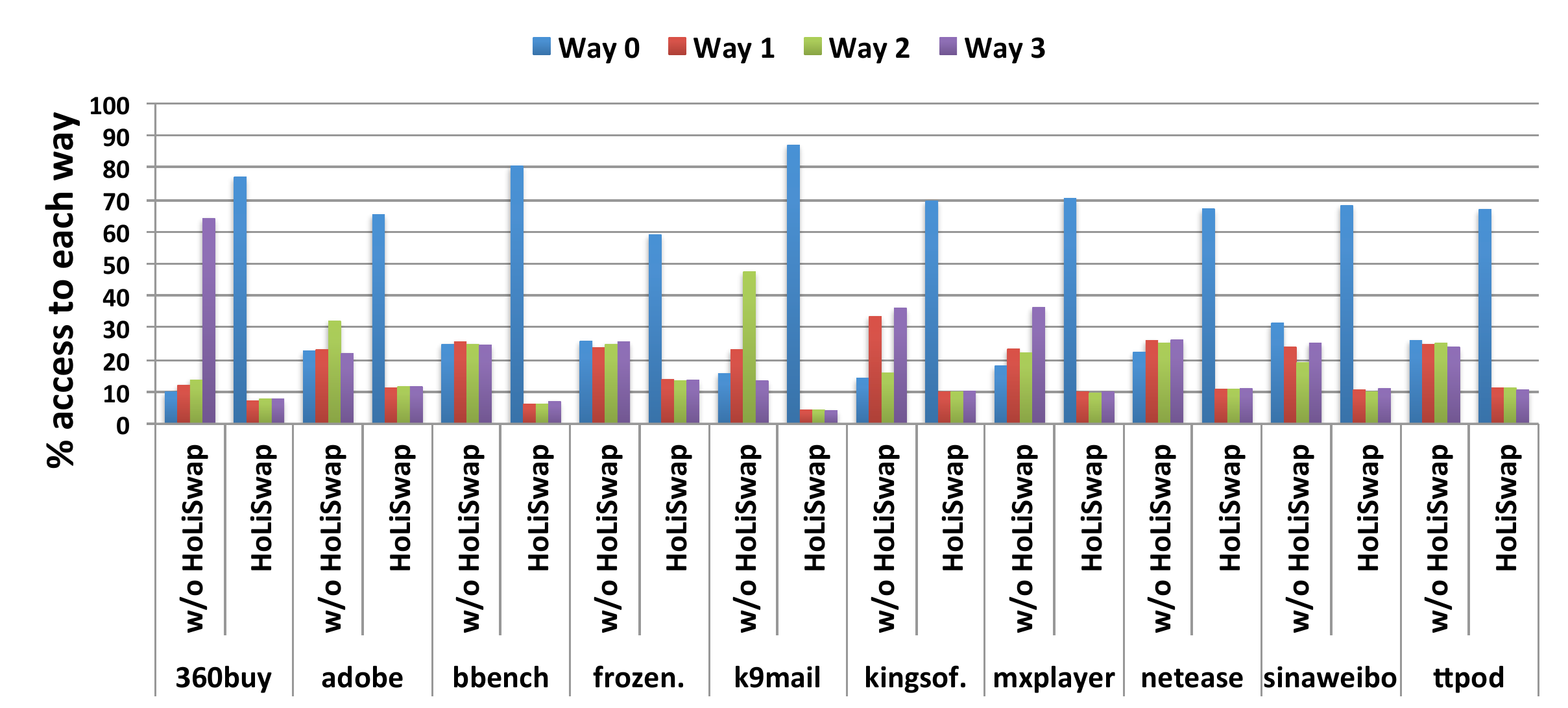}
\vspace{-10pt}
\caption{Fraction of accesses served by each way with and without HoLiSwap in Moby benchmarks.
}
\label{fi:way_access}
\end{figure}



Figure~\ref{fi:energy_perf} shows average performance and L1-D energy (normalized to {\sc Sequential} without HoLiSwap) for the 4 cache designs with and without HoLiSwap. L1-D energy is further broken down in SRAM array and wire components. 
HoLiSwap substantially reduces the L1-D cache wire energy for the first three designs.
The gains, as a fraction of the total L1-D cache energy, are largest for 
the {\sc Sequential} design (25.7\%) because it has the smallest SRAM array energy and most dominant wire component. 
HoLiSwap saves 41.1\% in the wire component for {\sc Sequential}.
HoLiSwap improves the energy of {\sc Parallel} by 12.1\% (44.0\% saving in wire energy component) of a substantially larger baseline. 
HoLiSwap provided 22.1\% energy improvement in L1-D cache with 44.6\% improvement in wire energy component in {\sc Prediction}. 
{\sc Filter} is independently more energy-efficient than the other schemes with HoLiSwap, as the L0 cache filters out most references to hot lines, but it has a performance loss of 2.9\%. HoLiSwap with {\sc Filter} provides 4.74\% energy savings over and above {\sc Filter}.
The energy savings include the overheads of swapping and logarithmic counters, which correspond to 0.15\% and 0.62\% of the L1-D cache energy, respectively. There is small performance degradation (\textless 0.13\%) due to blocking of cache ports during swaps in all cache designs, except for {\sc Prediction} which has 0.3\% performance drop. This is due to higher average memory latency in HoLiSwap as a result of lower way-prediction accuracy. 
Figure~\ref{fi:energy_savings} shows the L1-D energy saved by HoLiSwap on
each benchmark for each of the four designs.


Figure~\ref{fi:sweep} shows the sensitivity of L1-D cache energy and execution latency to epoch length $E$ for {\sc Sequential} with HoLiSwap (normalized to same scheme without HoLiSwap). The threshold $T$ was set to $E$/2. The energy savings range from 19.8\% ($E$=4) to 27.8\% ($E$=64) relative to the baseline. Small values of $E$ are impaired by redundant swaps while larger $E$ delays the identification of hot lines, resulting in lost opportunities for optimization. The performance degradation of HoLiSwap movement policy arising from blocking of cache port during cache line swap decreases with $E$, from 7.78\% for $E$=4 to 0.01\% for $E$=1024. The energy overhead of swaps supersede the energy saved at small epoch lengths ($E$=4) and with a heavy performance penalty, which explains why a policy like NuRAPID cannot be applied to L1-D cache. The overhead increases to 8.1\% when L1-I cache also uses HoLiSwap. At the same time, the out-of-order execution of the processor helps in hiding the additional memory latency during swaps to a great extent. Our choice of $E$=256 shows the best energy-delay product, also taking processor and DRAM energy into account, with an average of 0.13\% performance loss.


Figure~\ref{fi:way_pred} shows the way prediction accuracy for the proposed static way prediction, which always predicts $W_0$, provides 67.8\% accuracy on an average.
Our baseline way-predictor based on~\cite{powell2001reducing} had an accuracy of 81.8\% but with a higher implementation cost. Even with lower accuracy, static way-prediction along with HoLiSwap migration strategy provides 25.7\% energy improvement over baseline.

Energy improvements for various workloads are roughly correlated with the fraction of accesses to hot lines in Figure~\ref{fi:access}, but they also depend on the distribution of accesses to each way in absence of HoLiSwap migration policy. Figure~\ref{fi:way_access} shows the fraction of accesses served by each way with and without HoliSwap in L1-D cache. Largest energy improvements were observed in case of \textit{360buy} benchmark, since most of the accesses were made to the least energy-efficient way ($W_3$) of the cache (nearly all hot lines were placed in this way) by the replacement policy and relocated to the most energy-efficient way ($W_0$) by HoLiSwap. 



We also evaluated HoLiSwap for different L1-D cache sizes for {\sc Sequential}. The associativity of the cache (4-way) and the sizes of data subarrays (8KB) were kept unchanged. We changed the physical organisation in figure~\ref{fi:overview} to 2(rows)$\times$1(columns) and 2$\times$4 arrangement of subarrays for 16KB and 64KB cache, respectively. The energy range for accessing different ways decreased from 2.45$\times$ in 32KB cache to 1.54$\times$ in 16KB cache and with that, the energy savings from HoLiSwap also dropped from 27.8\% in 32KB cache to 9.5\% in 16KB cache. For 64KB cache, which had 3.16$\times$ energy variation in different ways, the energy savings increased to 34.3\%. The performance in 16KB cache was 0.71\% lower than that with 32KB cache, while 64KB cache improved the performance by 1.2\%. 

HoLiSwap, when applied to L1-I cache, results in 11.6\% energy saving with {\sc Parallel} scheme. Since performance is more sensitive to the L1-I cache latency, other lookup schemes, such as {\sc Sequential} (3.6\% performance drop), are rarely used. 

Using McPAT~\cite{li2009mcpat} on 22nm ARM A9 processor, we observed that the 32 KB L1-I cache with {\sc Parallel} corresponds to 8.6\% of the memory subsystem energy (caches + DRAM). The 32KB L1-D cache constitutes 14.7\% and 6.9\% of the memory subsystem energy using {\sc Parallel} and {\sc Sequential}, respectively. HoLiSwap can, therefore, save up to 3.42\% of the total memory subsystem energy and 1.82\% of total system energy (processor + DRAM) at 0.13\% combined performance loss.

\section{Related Work}



\begin{figure}
\vspace{-10pt}
\centering
\includegraphics[width=0.95\linewidth]{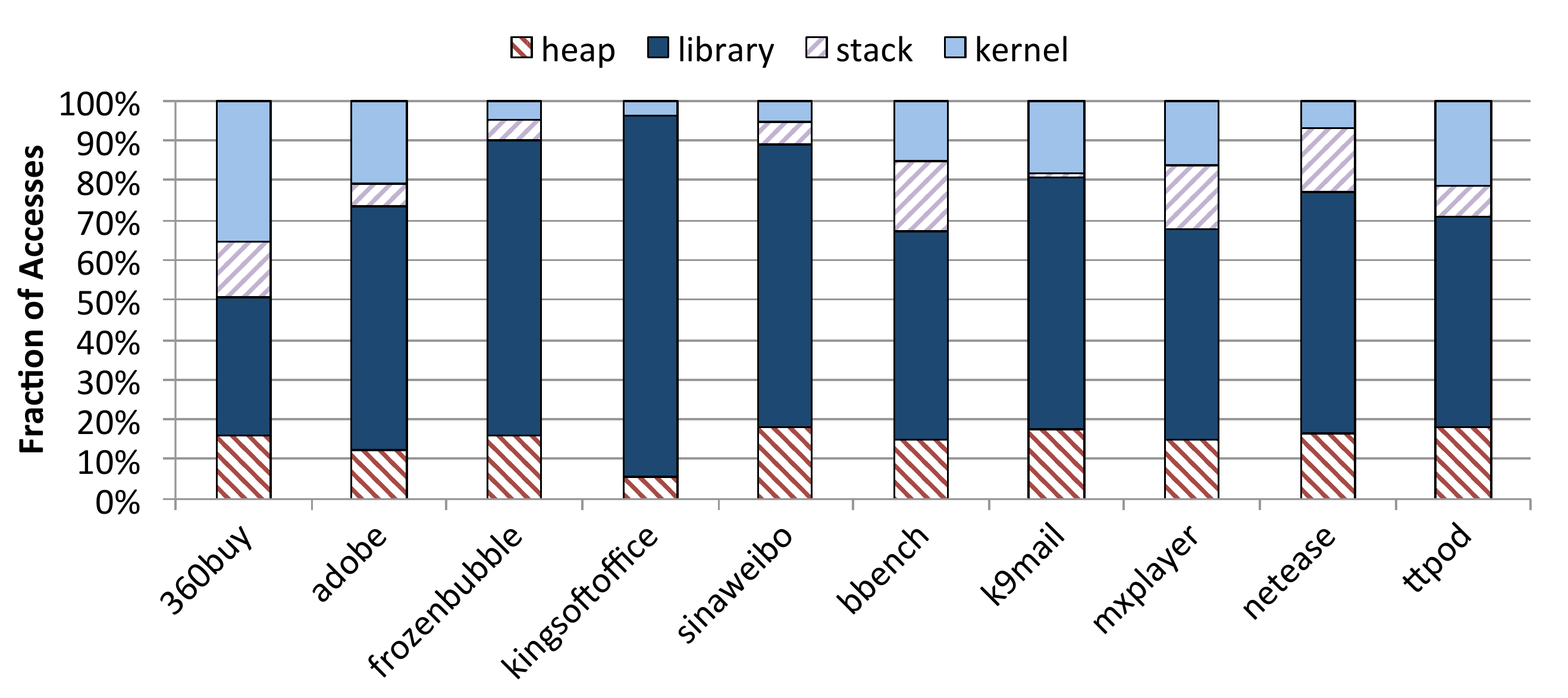}
\vspace{-15pt}
\caption{Fraction of references served from different address regions in Moby benchmarks.}
\label{fig:regionaccesses}
\end{figure}


Lee and Tyson~\cite{ref:regioncaching} observed that in SpecCPU2000 benchmarks~\cite{henning2000spec}, a majority of {\em hot} lines belonged to the stack region of the virtual address space (using a slightly different definition of {\em hot}). By using a small stack cache in addition to a large heap cache (L1-D cache), they achieved a power reduction of over 50\%.
However, the proposed region-based caching approach is not generally applicable to all workloads, such as mobile applications in Android. Figure~\ref{fig:regionaccesses} shows the relative distribution of accesses in the Android workloads from Moby benchmark suite belonging to the heap, shared library, stack and kernel regions. Since a number of objects get created dynamically in the heap space in these workloads and due to the presence of frequent system library calls, stack region only corresponds to only a small percentage of the total L1-D cache accesses. A majority of accesses belong to the heap and system library regions, making region-based caching approach ineffective. 
HoLiSwap overcomes this shortcoming by not differentiating between the hot lines belonging to the different regions of the cache. Besides, unlike region-based caching, HoLiSwap has no impact on the overall miss rate of the cache as it does not statically partition the cache.

HoLiSwap with {\sc Prediction} is similar to hash-rehash~\cite{agarwal1988cache}, column-associative~\cite{agarwal1993column} and multi-column caches~\cite{zhang1997two}, in which lines are swapped to a preferred way ($W_0$) that is also accessed first (often to reduce latency of tag lookup), but the main distinction of HoLiSwap is that it moves only the hot lines to $W_0$ at every epoch instead of the most recently used line after every reference.

\section{Conclusion}

In this paper we have described HoLiSwap which reduces L1 cache wire energy by swapping hot cache lines to the cache way nearest the processor.
This provides up to 44\% improvement in the wire energy with no impact on the cache miss rate. 
HoLiSwap can be employed with sequential or parallel cache access and can be combined with an L0 filter cache or cache way prediction.
When combined with cache way prediction, HoLiSwap simplifies the prediction by allowing the nearest way $W_0$ to always be predicted.
\singlespacing
\footnotesize
\bibliographystyle{IEEEtranS}
\bibliography{egbib}

\end {document}